\documentclass[twocolumn,showpacs,preprintnumbers,amsmath,amssymb]{revtex4}

\usepackage{graphicx}% Include figure files
\usepackage{dcolumn}% Align table columns on decimal point
\usepackage{bm}% bold math

\begin{document}

\title{Determinism and a supersymmetric classical model of quantum fields}% Force line breaks with \\

\author{H.-T. Elze}
\email{thomas@if.ufrj.br}

\affiliation{Dipartimento di Fisica, Via Filippo Buonarroti 2, I-56127 Pisa, Italia
\footnote{Permanent address: Instituto de F\'{\i}sica, Universidade Federal do Rio de Janeiro 
C.P. 68.528, 21941-972 Rio de Janeiro, RJ, Brazil}}

\date{February 25, 2005}% It is always \today, today,

\begin{abstract}
A {\it quantum} field theory is described which {\it is} a supersymmetric {\it classical} model. -- 
Supersymmetry generators of the system are used to split its Liouville operator into two contributions, 
with positive and negative spectrum, respectively. The unstable negative part is eliminated by a 
positivity constraint on physical states, which is invariant under the classical Hamiltonian flow. 
In this way, the classical Liouville equation becomes a functional Schr\"odinger equation of a genuine quantum 
field theory. Thus, 't\,Hooft's proposal to reconstruct quantum theory as emergent from an underlying 
deterministic system, is realized here for a field theory. Quantization is intimately related to the constraint, 
which selects the part of Hilbert space where the Hamilton operator is positive. This is seen as dynamical 
symmetry breaking in a suitably extended model, depending on a mass scale which discriminates classical 
dynamics beneath from emergent quantum mechanical behaviour.  
\end{abstract}


\maketitle

%%%%%%%%%%%%%%%%%%%%%%
\section{Introduction}
In a recent letter, I discussed anew the (dis)similarity between the classical Liouville 
equation and the Schr\"odinger equation \cite{I05}. In suitable coordinates both 
appear quite similar, apart from the characteristic doubling of the classical 
phase space degrees of freedom as compared to the quantum mechanical case. The Liouville 
operator is Hermitian in the operator approach to classical statistical mechanics 
developed by Koopman and von\,Neumann \cite{KN}. However, unlike the case of the 
quantum mechanical Hamiltonian, its spectrum is generally not bounded from below.  
Therefore, attempts to find a deterministic foundation of quantum theory -- based on a relation  
between the Koopman-von\,Neumann and quantum mechanical Hilbert spaces 
and equipped with the corresponding 
dynamics -- must particularly answer the question of how to construct a stable ground state.   
 
Investigations of these problems are to a large extent motivated by work of 
't\,Hooft, who has argued in favour of such model building, in order to gain 
a fresh look at the persistent clash between general relativity and quantum theory \cite{tHooft01}.  
Besides, since its very beginnings, there have been speculations 
about the possibility of deriving quantum theory from more fundamental and deterministic 
dynamical structures. The discourse running from Einstein, Podolsky and Rosen \cite{EPR} 
to Bell \cite{Bell}, and involving numerous successors, is well known, debating the (im)possibility of (local) hidden variables theories. 

Much of this debate has come under experimental scrutiny in recent years.  
No disagreement with quantum theory has been 
observed in the laboratory experiments on scales very large compared to the Planck scale.   
However, the feasible experiments cannot rule out the possibility that quantum mechanics 
emerges as an effective theory only on sufficiently large scales and can indeed be based 
on more fundamental models. 

In various examples, the emergence of a Hilbert space structure and unitary 
evolution in deterministic classical models has been demonstrated 
in an appropriate large-scale limit. However, in all cases, it is not trivial to assure 
that a resulting model qualifies as ``quantum'' by being built on a well-defined groundstate, 
i.e., with an energy spectrum that is bounded from below. 

A class of particularly simple emergent quantum models comprises systems which classically 
evolve in discrete time steps \cite{tHooft01,ES02}. Employing the path integral formulation of 
classical mechanics introduced by Gozzi and collaborators \cite{Gozzi}, it has been shown that 
actually a large class of classical models turns into unitary quantum mechanical ones, if the 
Liouville operator governing the statistical evolution is discretized \cite{I04}. However, 
there remains a large arbitrariness in such discretizations, which one would hope to 
reduce with the help of consistency or symmetry requirements of a more physical 
theory.  

Furthermore, it has been observed that classical systems with Hamiltonians which are linear 
in the momenta are also suitable for a reformulation in quantum mechanical terms. 
In order to provide a groundstate for such systems, 
a new kind of gauge fixing or constraints implementing 
``information loss'' at a fundamental level have been invoked  
\cite{tHooft01,Vitiello01,Blasone04}. Again, a unifying dynamical 
principle leading to the necessary truncation of the Hilbert space is still  
missing. 

Various other arguments for deterministically induced quantum features have been proposed 
recently -- 
see works collected in Part\,III of Ref.\,\cite{E04}, for example, or Refs.\,\cite{Smolin,Adler}, 
concerning statistical and/or dissipative systems, quantum gravity, and matrix models.   

Many of these attempts to base quantum theory on a classical footing can be seen as 
variants of the earlier stochastic quantization procedures of Nelson \cite{Nelson} and of 
Parisi and Wu \cite{Parisi}, often accompanied by a problematic analytic continuation from 
imaginary (Euclidean) to real time, in order to describe evolving systems. 

In distinction, one may aim at a truly dynamical understanding 
of the origin of quantum phenomena. In this work, I present a    
deterministic field theory from which a corresponding {\it quantum theory     
emerges by constraining the classical dynamics}. In particular, I will extend the 
globally supersymmetric (``pseudoclassical'') onedimensional model introduced in Ref.\,\cite{I05} 
to field theory. Thus, a functional Schr\"odinger equation is obtained with 
a positive Hamilton operator, involving the standard scalar boson part in the 
noninteracting case. 

Key ingredient is a splitting of the phase space evolution operator, i.e., of the 
classical Liouville operator, into positive and negative energy contributions. 
The latter, which would render the to-be-quantum field theory unstable, are   
eliminated by imposing a ``positivity constraint'' on the physical states, 
employing the Koopman-von\,Neumann approach \cite{KN,footnote1}. 
The splitting of the evolution operator and subsequent imposition 
of the constraint makes use of the supersymmetry of the classical system, which   
furnishes Noether charge densities which are essential here. 
While, technically, this is analogous to the imposition of the ``loss of information'' 
condition in 't\,Hooft's and subsequent work \cite{tHooft01,Vitiello01,Blasone04}, 
it is hoped that the present extension towards interacting fields opens a way 
to better understand the dynamical origin of such a constraint. While a  
dissipative information loss mechanism is plausible, alternatively a  
dynamical symmetry breaking may be considered as the cause. 

This paper is organized as follows. In Section\,II, the (pseudo)classical field 
theory is introduced and its equations of motion and global supersymmetry 
derived. Section\,III is devoted to the statistical mechanics of an ensemble 
of such systems, its Hilbert space description and Liouville equation, in particular. 
The Liouville equation 
is then cast into the form of a functional Schr\"odinger equation in Section\,IV.  
Also the necessary positivity constraint on physical states is discussed, constructed, and incorporated 
there which turns the 
emergent Hamiltonian into a positive local operator with a proper 
quantum mechanical groundstate.  
In the concluding Section\,V, I mention some interesting 
topics for further exploration, especially the relation of the positivity constraint 
to symmetry breaking.  

%%%%%%%%%%%%%%%%%%%%%%%%%%%%%%%%%%%%%%%%%%%%
\section{The supersymmetric classical model}
The following derivation will newly make use of ``pseudoclassical mechanics'' or, rather,  
pseudoclassical field theory. These notions have been 
introduced through the work of Casalbuoni and of Berezin and 
Marinov, who considered a {\it Grassmann variant of classical mechanics}, 
studying the dynamics of spin degrees of freedom classically and 
after quantization in the usual way \cite{CB}. 

Classical mechanics based on Grassmann algebras has more recently found much attention 
in attempts to better understand the zerodimensional limit of classical and 
quantized supersymmetric field theories, see Refs.\,\cite{FDeW,MJ} 
and further references therein.  

Let us introduce a ``fermionic'' field $\psi$, together with a real scalar field $\phi$. 
The former is represented by the nilpotent generators of an infinite dimensional 
Grassmann algebra \cite{footnote2}.   
They obey:  
\begin{equation}\label{odd} 
\{ \psi (x),\psi (x')\}_+\equiv\psi (x)\psi (x')+\psi (x')\psi (x)=0 
\;\;, \end{equation}
where $x,x'$ are coordinate labels in Minkowski space. All elements are real. 

Then, the classical model to be studied is defined by the action: 
\begin{equation}\label{action}
S\equiv\int\mbox{d}^4x\;\Big (\dot\phi\dot\psi -\phi\big (-\Delta +m^2+v(\phi)\big )\psi\Big )\equiv\int\mbox{d}t\;L 
\;\;, \end{equation} 
where dots denote time derivatives, and $v(\phi )$ may be a polynomial in $\phi$, for example. 

This system apparently has not been studied before, which might be related to the fact that the action 
is Grassmann odd. However, in line with the present attempt to find a    
classical foundation of a quantum field theory, no path integral quantization (or other) of the 
model is intended, which could be obstructed by such a fermionic action.  
  
Introducing canonical momenta, 
\begin{equation}\label{Pphipsi}
P_\phi\equiv\frac{\delta L}{\delta\dot\phi }=\dot\psi\;\;, 
\;\;\;P_\psi\equiv\frac{\delta L}{\delta\dot\psi }=\dot\phi
\;\;, \end{equation} 
as usual, one calculates the Hamiltonian,  
\begin{eqnarray}
H&=&\int\mbox{d}^3x\;\Big (P_\phi\dot\phi+P_\psi\dot\psi\Big )-L 
\nonumber \\ [1ex] \label{Hamiltonian} 
&=&\int\mbox{d}^3x\;\Big (P_\phi P_\psi +\phi K\psi\Big )  
\;\;, \end{eqnarray} 
which turns out to be Grassmann odd as well. Here the first of two 
useful abbreviations has been introduced: 
$K\equiv -\Delta +m^2+v(\phi )$, $K'\equiv K+\phi\mbox{d}v(\phi )/\mbox{d}\phi$.   
   
Hamilton's equations of motion for our model follow:   
\begin{eqnarray}\label{pphi}
\dot\phi &=&\frac{\delta H}{\delta P_\phi }=P_\psi
\;\;, \\ [1ex] \label{ppsi}
\dot\psi &=&\frac{\delta H}{\delta P_\psi }=P_\phi
\;\;, \\ [1ex] \label{pphidot}  
\dot P_\phi &=&-\frac{\delta H}{\delta\phi }=-K'\psi
\;\;, \\ [1ex] \label{ppsidot}
\dot P_\psi &=&-\frac{\delta H}{\delta\psi }=-K\phi 
\;\;. \end{eqnarray} 
Combining the equations, one obtains: 
\begin{equation}\label{eom} 
\ddot\phi =-K\phi\;\;,\;\;\;\ddot\psi =-K'\psi 
\;\;, \end{equation} 
i.e., the generally nonlinear field equations, where there is only a parametric coupling 
between the fields $\phi$ and $\psi$, namely of the former 
to the latter.  

These equations are invariant under 
the global symmetry transformation, 
\begin{equation}\label{sym1} 
\phi\longrightarrow\phi +\epsilon\psi   
\;\;, \end{equation} 
where $\epsilon$ is an infinitesimal real parameter. Associated is the Noether charge:  
\begin{equation}\label{C1} 
C_1\equiv\int\mbox{d}^3x\;P_\phi\psi 
\;\;, \end{equation} 
which is a constant of motion. Similarly, a second global symmetry transformation 
leaves the system invariant: 
\begin{equation}\label{sym2} 
\psi\longrightarrow\psi +\epsilon\dot\phi   
\;\;, \end{equation} 
with associated conserved Noether charge: 
\begin{equation}\label{C2}
C_2\equiv\int\mbox{d}^3x\;\Big (\frac{1}{2}P_\psi^2+V(\phi )\Big )  
\;\;, \end{equation} 
which is the total energy of the classical scalar 
field, with $\mbox{d}V(\phi )/\mbox{d}\phi\equiv K\phi$, appropriately taking care of  
gradient terms by partial integration.  

In the following, it will be useful to introduce the Poisson bracket operation acting on two 
observables $A$ and $B$, which generally can be function(al)s of the phase space variables 
$\phi,P_\phi,\psi,P_\psi$: 
\begin{eqnarray}
\{A,B\}&\equiv&A\int\mbox{d}^3x\;\Big (     
\frac{\stackrel{\leftharpoonup}{\delta}}{\delta P_\phi}\frac{\stackrel{\rightharpoonup}{\delta}}{\delta\phi}
+\frac{\stackrel{\leftharpoonup}{\delta}}{\delta P_\psi}\frac{\stackrel{\rightharpoonup}{\delta}}{\delta\psi}
\nonumber \\ [1ex] \label{PB} 
&\;&\;\;\;\;\;\;\;\;\;\;\;\;\;\;
-\frac{\stackrel{\leftharpoonup}{\delta}}{\delta\phi}\frac{\stackrel{\rightharpoonup}{\delta}}{\delta P_\phi}
-\frac{\stackrel{\leftharpoonup}{\delta}}{\delta\psi}\frac{\stackrel{\rightharpoonup}{\delta}}{\delta P_\psi}
\Big )B  
\;\;, \end{eqnarray} 
where all functional derivatives refer to the same space-time argument and act in the 
indicated direction; for the fermionic variables this  
direction is meant to coincide with their left/right-derivative character \cite{FDeW}. 

Note that $\{ A,B\}=-\{ B,A\}$, if the derivatives of $A$ and $B$ 
commute, i.e., if in each contributing term at least one of the two is Grassmann even. 
Furthermore, for any observable $A$, the usual relation among time derivatives holds: 
\begin{equation}\label{timederivs} 
\frac{\mbox{d}}{\mbox{d}t}A=\{ H,A\}+\partial_tA
\;\;, \end{equation}
which embodies Hamilton's equations of motion.  

Naturally, the time independent Hamiltonian of Eq.\,(\ref{Hamiltonian}) is conserved by the evolution 
according to the classical equations of motion. 

For the Hamiltonian and Noether charge densities, identified by  
$H\equiv\int\mbox{d}^3xH(x)$ and $C_j\equiv\int\mbox{d}^3xC_j(x)|_{j=1,2}$, 
respectively, one finds a local (equal-time) supersymmetry algebra:  
\begin{equation}\label{SUSY1} 
\{ C_1(x),C_2(x')\}=-H(x)\delta^3(x-x') 
\;\;, \end{equation} 
and, 
\begin{eqnarray} 
\{ H(x),C_1(x')\}+\{ C_1(x'),H(x)\}&\;&
\nonumber \\ [1ex] \label{SUSY2} 
=\;\{ H(x),C_1(x')\}+\{ C_1(x),H(x')\}&=&0\;\;,
\\ [1ex] \label{SUSY3} 
\{ H(x),C_2(x')\}&=&0\;\;,   
\\ [1ex] \label{SUSY4}
\{ C_j(x),C_j(x')\}\;=\;\{ H(x),H(x')\}&=&0 
\;\;. \end{eqnarray}
In all calculations, eventually arising coincidence limits are assumed to be smooth, 
since classical fields are involved. 
Of course, for any one of the constants of motion, 
$A\in\{ H,C_1,C_2\}$, one obtains: $\{ H,A\}=\dot A=0$.  

In the following section, the present analysis is applied to the 
corresponding phase space representation of an ensemble of systems  
and, furthermore, developed into an equivalent Hilbert space picture. 

%%%%%%%%%%%%%%%%%%%%%%%%%%%%%%%%%%%%%%%%%%%%%%%%%%%%%%%%%
\section{The Liouville equation: from the field theory in 
phase space to the Hilbert space picture}  
A particular example of Eq.\,(\ref{timederivs}) is the Liouville equation 
for a conservative system, such as the model considered in Section\,II.  
Considering an ensemble of systems, especially with some distribution over different initial conditions, 
this equation governs the evolution of its phase space density $\rho$: 
\begin{equation}\label{Liouville} 
0=i\frac{\mbox{d}}{\mbox{d}t}\rho =i\partial_t\rho -\hat{\cal L}\rho
\;\;, \end{equation}
where a convenient factor $i$ has been introduced, and the Liouville operator 
$\hat{\cal L}$ is defined by: 
\begin{equation}\label{Liouvilleop}  
-\hat{\cal L}\rho\equiv i\{ H,\rho \}
\;\;. \end{equation}
These equations summarize the classical statistical mechanics of a conservative system, 
given the Hamiltonian $H$ in terms of the phase space variables.  

Next, let us briefly recall the equivalent Hilbert space formulation 
developed by Koopman and von\,Neumann \cite{KN}. It will be modified here 
in a way appropriate for the supersymmetric classical field theory in question.  

Two postulates are put forth: \begin{itemize} 
\item (A) the phase space density functional can be factorized in the form $\rho\equiv\Psi^*\Psi$;
\item (B) the Grassmann valued and, in general, complex state functional $\Psi$ itself obeys the 
Liouville Eq.\,(\ref{Liouville}). 
\end{itemize} 
Furthermore, the complex valued inner product of such state functionals is defined by: 
\begin{equation}\label{scalarprod}
\langle\Psi |\Phi\rangle\equiv\int {\cal D}\phi {\cal D}P_\psi {\cal D}\psi {\cal D}P_\phi\; 
\Psi^*\Phi =\langle\Phi |\Psi\rangle^*
\;\;, \end{equation}
i.e., by functional integration over all phase space variables (fields). 
However, due to the presence of Grassmann valued variables, the $*$-operation which defines  
the dual of a state functional needs 
special attention and will be discussed shortly.

The above definitions make sense for functionals which suitably generalize the notion of 
square-integrable functions. In particular, the   
functional integrals can be treated rigorously by discretizing the system, properly   
pairing degrees of freedom.

Given the Hilbert space structure, the Liouville operator of a conservative system has to be 
Hermitian and the overlap $\langle\Psi |\Psi\rangle$ is a conserved quantity. 
Then, the Liouville equation also applies to 
$\rho =|\Psi |^2$, due to its linearity, and $\rho$ may be  
interpreted as a probability density, as before \cite{KN}.   
Naturally, this is needed for meaningful phase space expectation values of observables. 

Certainly, one is reminded here of the usual quantum mechanical formalism. In order to 
expose the striking similarity as well as the remaining crucial difference, further transformations of the functional Liouville equation are useful \cite{I05}.   

A Fourier transformation replaces the momentum $P_\psi$ by a second scalar field      
$\bar\phi$. Furthermore, define $\bar\psi\equiv P_\phi$.  
Thus, the Eqs.\,(\ref{Liouville})--(\ref{Liouvilleop}) yield: 
\begin{equation}\label{Schroedinger} 
i\partial_t\Psi =\hat{\cal H}\Psi
\;\;, \end{equation} 
where $\Psi$ is considered as a functional of $\phi,\bar\phi,\psi,\bar\psi$, 
and with the {\it emergent} ``Hamilton operator'': 
\begin{eqnarray}\label{Hem}
&\;&\hat{\cal H}\Psi\equiv -i\int {\cal D}P_\psi \;\exp (iP_\psi\cdot\bar\phi )\{ H,\Psi\} 
\\ [1ex] \nonumber 
&=&\int\mbox{d}^3x\;\Big (-\delta_{\bar\phi}\delta_\phi 
+\bar\phi K\phi 
-i(\bar\psi\delta_\psi -\psi K'\delta_{\bar\psi})\Big )\Psi
\\ [1ex] \label{Hem1}
&\equiv&\int\mbox{d}^3x\;\hat{\cal H}(x)\;\Psi
\;\;, \end{eqnarray}
using the abbreviation $f\cdot g\equiv\int\mbox{d}^3x\;f(x)g(x)$. Note that the 
density $\hat{\cal H}(x)$ is Grassmann even.

While the Eq.\,(\ref{Schroedinger}) strongly resembles a functional Schr\"odinger equation, 
several comments must be made here which point out its  
different character. 

First of all, following a linear transformation of the 
scalar field variables, $\phi\equiv (\sigma +\kappa )/\sqrt 2$ and 
$\bar\phi\equiv (\sigma -\kappa )/\sqrt 2$, one finds a ``bosonic''  
kinetic energy term: 
\begin{equation}\nonumber 
-\frac{1}{2}\int\mbox{d}^3x\;\left (\delta_\sigma^{\;2}-\delta_\kappa^{\;2}\right ) 
\;\;, \end{equation} 
which is {\it not bounded from below}. Therefore, neglecting the Grassmann variables momentarily, 
the remaining Hermitian part of the Hamiltonian lacks a lowest energy state, which otherwise could 
qualify as the emergent quantum mechanical groundstate of the bosonic sector.   

Secondly, as could be expected, the fermionic sector reveals a similar problem. 

The $*$-operation mentioned before amounts to complex conjugation for a bosonic 
state functional, $(\Psi [\bar\phi ,\phi ])^*\equiv\Psi^*[\bar\phi ,\phi ]$, analogously to    
an ordinary wave function in quantum mechanics. However, based on complex conjugation alone, 
the fermionic part of the Hamiltonian (\ref{Hem}) would not be Hermitian. 

Instead, a detailed construction of the inner 
product for functionals of Grassmann valued fields has been presented in 
Ref.\,\cite{Jackiw}; see also further examples in Refs.\,\cite{KieferRoskies}. 
Considering only the {\it noninteracting case} 
with $K'=K$, i.e., with $v(\phi )=0$ in Eq.\,(\ref{action}), the construction of   
Floreanini and Jackiw can be directly applied here. Then, the Hermitian 
conjugate of $\psi$ is $\psi^\dagger =\delta_\psi$ and of $\bar\psi$ it is $\bar\psi^\dagger =\delta_{\bar\psi}$. 
Furthermore, rescaling $\bar\psi\;\longrightarrow\;\bar\psi\sqrt K$,    
the fields $\bar\psi$ and $\psi$ obtain the same dimensionality. Together, this  
suffices to render Hermitian the fermionic part of the Hamiltonian (\ref{Hem}), 
which becomes: 
\begin{equation}\label{HemF} 
\hat{\cal H}_{\bar\psi\psi}\equiv i(\psi\sqrt K\delta_{\bar\psi} -\bar\psi\sqrt K\delta_\psi )
\;\;. \end{equation}  
In the presence of interactions, with $K'\neq K$, additional modifications are 
necessary and will be considered elsewhere. In any case, although $\hat{\cal H}_{\bar\psi\psi}$ 
must be (made) Hermitian, its eigenvalues generally will not have a lower bound either. 

To summarize, the emergent Hamiltonian $\hat{\cal H}$ 
tends to be unbounded from below, thus lacking a groundstate. 
This generic difficulty has been encountered in various attempts to build deterministic 
quantum models, i.e., classical models which can simultaneously be seen as 
quantum mechanical ones \cite{tHooft01,ES02,I04,Vitiello01,Blasone04}. 
For the present case, this will be discussed and resolved in Section\,IV.    

To conclude this section, equal-time operator relations for the interacting case are derived here, 
which are related to the supersymmetry algebra 
of Eqs.\,(\ref{SUSY1})--(\ref{SUSY4}). This is achieved by Fourier 
transformation of appropriate Poisson brackets, similarly as with the emergent Hamiltonian 
in Eq.\,(\ref{Hem}) above. 

To begin with, the operators corresponding to the Noether 
densities will be useful. Using Eq.\,(\ref{C1}) and $\bar\psi\equiv P_\phi$, as before, 
one obtains:  
\begin{eqnarray} 
\hat{\cal C}_1(x)\Psi&\equiv&\int {\cal D}P_\psi \;\exp (iP_\psi\cdot\bar\phi )\{ C_1(x),\Psi\}
\nonumber \\ [1ex] \label{C1op}
&=&\big (-\psi\delta_\phi +i\bar\psi\bar\phi\big )_{(x)}\Psi 
\;\;. \end{eqnarray} 
Similarly, one obtains:  
\begin{equation}\label{C2op}
\hat{\cal C}_2(x)\Psi\;\equiv\;
\big (-i\delta_{\bar\phi}\delta_\psi -\phi K\delta_{\bar\psi}\big )_{(x)}\Psi
\;\;, \end{equation} 
which is related to Eq.\,(\ref{C2}). 

Both operators are Grassmann odd and obey: 
\begin{equation}\label{Canticomm} 
\{ \hat{\cal C}_j(x),\hat{\cal C}_j(x')\}_+=0
\;\;, \end{equation}
for $j=1,2$. Therefore, they are nilpotent, $\hat{\cal C}_j^{\;2}(x)=0$. 
This should be compared to Eq.\,(\ref{SUSY4}), as well as 
the vanishing commutator: 
\begin{equation}\label{Hcomm} 
[\hat{\cal H}(x),\hat{\cal H}(x')]=0 
\;\;, \end{equation} 
where $[\hat A,\hat B]\equiv\hat A\hat B-\hat B\hat A$. Thus, 
the emergent theory is local, as expected. 

It should be remarked that in all calculations of (anti)commutation relations 
eventually necessary partial integrations, i.e. shifting of gradients, are 
justified by smearing with suitable test functions and integrating.  

Further relations that correspond to Jacobi identities on the level 
of the Poisson brackets are interesting. Generally, one has to be careful about extra signs that arise 
due to the Grassmann valued quantities, as compared to more familiar ones related 
to real or complex variables \cite{FDeW}. 
Straightforward calculation gives: 
\begin{eqnarray}\label{SUSYop1} 
[\hat{\cal H}(x),\hat{\cal C}_j(x')]&=&0 
\;\;,\;\;\mbox{for}\;j=1,2 
\;\;, \\ [1ex] \label{SUSYop2}
\{ i\hat{\cal C}_1(x),\hat{\cal C}_2(x')\}_+&=&\hat{\cal H}(x)\delta^3(x-x')
\;\;, \end{eqnarray} 
cf. Eqs.\,(\ref{SUSY1})--(\ref{SUSY3}); the extra factor $i$ must be attributed to the Fourier transformation 
that enters between the phase space functions before and the operators here.  

Finally, it is noteworthy that a copy of the above operator algebra arises, 
if one performs the replacements $\psi\leftrightarrow -\delta_{\bar\psi}$ and 
$\bar\psi\leftrightarrow\delta_\psi$ on the operators $\hat{\cal C}_j$. This yields the nilpotent operators 
$\hat{\cal D}_j$, instead of the $\hat{\cal C}_j$: 
\begin{eqnarray}\label{D1} 
i\hat{\cal D}_1(x)&\equiv&\big (i\delta_{\bar\psi}\delta_\phi -\delta_\psi\bar\phi\big )_{(x)}
\;\;, \\ [1ex] \label{D2} 
\hat{\cal D}_2(x)&\equiv&\big (i\delta_{\bar\phi}\bar\psi -\phi K\psi\big )_{(x)}
\;\;, \end{eqnarray}
with a convenient overall sign introduced in the latter definition. They fullfill the same 
(anti)commutation relations as in Eqs.\,(\ref{Canticomm})--(\ref{SUSYop2}). 

Finally, also the following local operators commute with the Hamiltonian density: 
\begin{equation}\label{CD} 
(i\hat{\cal D}_1\hat{\cal C}_2\pm i\hat{\cal C}_1\hat{\cal D}_2) 
=-i(\delta_{\bar\psi}\delta_\psi\mp\bar\psi\psi )(-\delta_{\bar\phi}\delta_\phi +\bar\phi K\phi )
\;\;, \end{equation}
with $[\hat{\cal C}_1,\hat{\cal D}_2]=[\hat{\cal D}_1,\hat{\cal C}_2]=0$. These 
operators are not nilpotent. Their square, though, is highly singular. 

One may complete these considerations with the full set of  
operators generating the ordinary space-time symmetries of our model. 
However, they are not believed to play a special role for the considerations 
of the following section. There, the no-groundstate problem of the emergent Hamiltonian, 
Eq.\,(\ref{Hem}), will be addressed.  

%%%%%%%%%%%%%%%%%%%%%%%%%%%%%%%%%%%%%%%%%%%%%%%%%%%%%%%%%%%%%%%%%
\section{Providing the groundstate of the emergent quantum model}
Following Eq.\,(\ref{Hem}), it has been pointed out that the emergent Hamiltonian 
lacks a proper groundstate, i.e., its spectrum is not bounded from below. This 
prohibits to interpret the model, as it stands, as a quantum mechanical one already, 
despite close formal similarities. 

In order to overcome this difficulty, the general strategy is to find a positive 
definite local operator $\hat P$ that commutes with the Hamiltonian density, 
$[\hat{\cal H}(x),\hat P(x')]=0$. Then, the Hamiltonian can be split into 
contributions with positive and negative spectrum: 
\begin{equation}\label{Hemsplit} 
\hat{\cal H}=\hat{\cal H}_+-\hat{\cal H}_-
\;\;, \end{equation}
where: 
\begin{equation}\label{Hempm}
\hat{\cal H}_\pm
\equiv\int\mbox{d}^3x\;F\big (\hat{\cal H}(x)\pm\hat P(x)\big )
\;\;. \end{equation} 
Here $F$ can be any even function with the property: 
\begin{equation}\label{F} 
F(a+b)-F(a-b)=abG(a^2,b^2)\;\;,\;\;G>0 
\;\;, \end{equation} 
for $a,b\in\mathbf{R}$. 

The simplest example is $F(a)\equiv a^2,\;G\equiv 4$. With this,  
the splitting of $\hat{\cal H}$ is explicitly given by: 
\begin{equation}\label{Hemsplit1} 
\hat{\cal H}=\int\mbox{d}^3x\;\Big (\frac{(\hat{\cal H}+\hat P)^2-(\hat{\cal H}-\hat P)^2}{4\hat P}\Big )
\;\;, \end{equation} 
i.e., $\hat{\cal H}_\pm (x)=(\hat{\cal H}(x)\pm\hat P(x))^2/4\hat P(x)$. 
A quartic polynomial could be used instead, etc. In the absence 
of further symmetry requirements, or other, from the model under consideration, the 
simplest splitting will do. It will allow us to obtain a free quantum field 
theory, in particular, as leading part of the relevant Hamilton operator. 

Here, as in the following, a regularization is necessary, in order to give a 
meaning particularly to some of the squared operators that will keep appearing. 

Finally, the spectrum of the Hamiltonian $\hat{\cal H}$ is made 
bounded from below by imposing the ``positivity constraint'': 
\begin{equation}\label{constraint} 
\hat{\cal H}_-\Psi =0 
\;\;. \end{equation}
This constraint can be enforced as an initial condition, for example, and is 
preserved by the evolution, since $[\hat{\cal H}_+(x),\hat{\cal H}_-(x)]=0$,  
by construction. In this way, the {\it physical states} of the system are selected  
which are based on the existence of a {\it quantum mechanical groundstate}. 

Such a constraint selecting the physical part of the emergent Hilbert space has been 
earlier discussed in the models of Refs.\,\cite{tHooft01,Vitiello01,Blasone04}. 
It has been interpreted by 't\,Hooft as ``information loss'' at the fundamental level where 
quantum mechanics may arise from a deterministic theory. However, it seems also quite 
possible to relate this to a {\it dynamical symmetry breaking} phenomenon instead, 
cf. Section\,V.   

For our field theory, the noninteracting and interacting cases 
shall now be studied separately in more detail.

%%%%%%%%%%%%%%%%%%%%%%%%%%%%%%%%%%%%
\subsection{The noninteracting case}
As mentioned before, with $v(\phi )=0$ in Eq.\,(\ref{action}), and therefore $K'=K=-\Delta +m^2$, 
the rescaling $\bar\psi\;\longrightarrow\;\bar\psi\sqrt K$ is useful, and one may 
consider the set of operators: 
\begin{eqnarray}\label{Hemresc}  
\hat{\cal H}(x)&=&\big (-\delta_{\bar\phi}\delta_\phi +\bar\phi K\phi \big )_{(x)}
+\hat{\cal H}_{\bar\psi\psi}(x)
\;\;, \\ [1ex] \label{C1resc}
i\hat{\cal C}_1(x)&=&
\big (-i\psi\delta_\phi -\bar\psi\sqrt K \bar\phi\big )_{(x)}
\;\;, \\ [1ex] \label{C2resc}
\hat{\cal C}_2(x)&=&
\big (-i\delta_{\bar\phi}\delta_\psi -\phi\sqrt K \delta_{\bar\psi}\big )_{(x)}
\;\;, \end{eqnarray} 
with $\hat{\cal H}_{\bar\psi\psi}$ from Eq.\,(\ref{HemF}). These operators 
fullfill the same operator algebra as discussed in the previous section. 

Furthermore, let us consider the Hermitian conjugate operators, in this case 
based on $\psi^\dagger =\delta_\psi$ and 
$\bar\psi^\dagger =\delta_{\bar\psi}$ \cite{Jackiw}: 
\begin{eqnarray}\label{C1adj}
\big (i\hat{\cal C}_1(x)\big )^\dagger&=&
\big (-i\delta_\psi\delta_\phi -\delta_{\bar\psi}\sqrt K \bar\phi\big )_{(x)}
\;\;, \\ [1ex] \label{C2adj}
\big (\hat{\cal C}_2(x)\big )^\dagger&=&
\big (-i\delta_{\bar\phi}\psi -\phi\sqrt K\bar\psi\big )_{(x)}
\;\;. \end{eqnarray}  
They commute with the Hermitian density $\hat{\cal H}(x)$, and one finds that  
$\{ i\hat{\cal C}_1(x),(\hat{\cal C}_2(x))^\dagger\}_+=0$, together with the corresponding  
adjoint relation. 

Then, also the following Hermitian operators commute with the Hamiltonian density: 
\begin{eqnarray} 
&\;&\hat{\cal C}_{1+}(x)\;\equiv\;\big (i\hat{\cal C}_1(x)+\big (i\hat{\cal C}_1(x)\big )^\dagger\big )/\sqrt 2 
\nonumber \\ [1ex] \label{C1+}
&\;&=\frac{1}{\sqrt 2}\big (-i(\delta_\psi +\psi )\delta_\phi -(\delta_{\bar\psi}+\bar\psi )\sqrt K \bar\phi\big )_{(x)}
\;, \\ [1ex]  
&\;&\hat{\cal C}_{2+}(x)\;\equiv\;\big (\hat{\cal C}_2(x)+\big (\hat{\cal C}_2(x)\big )^\dagger\big )/\sqrt 2 
\nonumber \\ [1ex] \label{C2+}
&\;&=\frac{1}{\sqrt 2}\big (-i(\delta_\psi +\psi )\delta_{\bar\phi}-(\delta_{\bar\psi}+\bar\psi )\sqrt K\phi\big )_{(x)}
\;. \end{eqnarray}
These operators are interesting, since they present, in some sense, the 
{\it ``square-root of the harmonic oscillator''}: 
\begin{eqnarray}\label{C1+2}
\hat{\cal C}_{1+}^{\;2}(x)&=& 
\frac{\delta^3(0)}{2}\big (-\delta_\phi^{\;2}+\bar\phi K\bar\phi\big )_{(x)}
\;\;, \\ [1ex] \label{C2+2}
\hat{\cal C}_{2+}^{\;2}(x)&=&
\frac{\delta^3(0)}{2}\big (-\delta_{\bar\phi}^{\;2}+\phi K\phi\big )_{(x)} 
\;\;, \end{eqnarray}
or, rather, since the sum of the squared operators amounts to the Hamiltonian density of two free bosonic quantum fields.

It seems natural now to choose the positive definite local operator $\hat P$ of 
Eq.\,(\ref{Hemsplit1}) as: 
\begin{equation}\label{Pfree} 
\hat P(x)\equiv\frac{\xi}{\delta^3(0)}\big (\hat{\cal C}_{1+}^{\;2}(x)+\hat{\cal C}_{2+}^{\;2}(x)\big ) 
\;\;, \end{equation}
where $\xi$ is a dimensionless parameter.  
This results in the operators of definite sign: 
\begin{eqnarray}\label{Hemplus} 
\hat{\cal H}_\pm (x)&=&\big (\hat{\cal H}(x)\pm\hat P(x)\big )^2/4\hat P(x) 
\nonumber \\ [1ex] 
&=&\frac{\xi}{8}\big (-\delta_\phi^{\;2}+\phi K\phi -\delta_{\bar\phi}^{\;2}+\bar\phi K\bar\phi\big )  
\nonumber \\ [1ex] \label{Hemplus1} 
&\;&\pm\frac{1}{2}\hat{\cal H}(x)+\frac{1}{4}\hat{\cal H}^2(x)/\hat P(x) 
\;\;, \end{eqnarray}
cf. Eqs.\,(\ref{Hemsplit})--(\ref{Hemsplit1}). 

Setting $\xi =2$ and performing again the linear transformation $\phi\equiv (\sigma +\kappa )/\sqrt 2$ and 
$\bar\phi\equiv (\sigma -\kappa )/\sqrt 2$, previously mentioned after Eqs.\,(\ref{Hem})--(\ref{Hem1}), here instead yields the Hamiltonian density: 
\begin{equation}\label{Hemplus2}
\hat{\cal H}_+(x)=\frac{1}{2}\Big (-\delta_\sigma^{\;2}+\sigma K\sigma 
+\hat{\cal H}_{\bar\psi\psi}+\frac{1}{2}\hat{\cal H}^2/\hat P\Big )_{(x)} 
\;\;, \end{equation} 
with $\hat{\cal H}_{\bar\psi\psi}$ from Eq.\,(\ref{HemF}), and 
where, of course, the linear transformation has also been performed in $\hat{\cal H}^2/\hat P$. 
One observes that the only trace of the previous instability is now relegated to this last   
term, which still involves the scalar field $\kappa$. The {\it local interactions} 
present in this term certainly have a nonstandard form. Additional parameters  
playing the role of coupling constants could be introduced by a more complicated 
splitting of the emergent Hamiltonian, see Eqs.\,(\ref{Hemsplit})--(\ref{Hemsplit1}), or a different 
choice for the operator $\hat P$.

However, the Hamilton operator $\hat{\cal H}_+$ 
has a positive spectrum, by construction, and the leading  
terms 
are those of a {\it free bosonic quantum field} together with a {\it fermion doublet}  
in the Schr\"odinger representation. They dominate at low energy.   

Similarly, the constraint operator density becomes: 
\begin{equation}\label{Hemmin}
\hat{\cal H}_-(x)=\frac{1}{2}\Big (-\delta_\kappa^{\;2}+\kappa K\kappa  
-\hat{\cal H}_{\bar\psi\psi}+\frac{1}{2}\hat{\cal H}^2/\hat P\Big )_{(x)} 
\;\;. \end{equation} 
A certain symmetry with Eq.\,(\ref{Hemplus2}) is obvious; note that 
$-\hat{\cal H}_{\bar\psi\psi}=\hat{\cal H}_{\psi\bar\psi}$. It suggests to think of the elimination 
of part of the Hilbert space, Eq.\,(\ref{constraint}), as a 
dynamical symmetry breaking effect. This point will be briefly addressed 
in the concluding section.
  
%%%%%%%%%%%%%%%%%%%%%%%%%%%%%%%%%
\subsection{The interacting case}
In the interacting case, one has $v(\phi )\neq 0$ in Eq.\,(\ref{action}), 
$K\equiv -\Delta +m^2+v(\phi )$, and  
$K'\equiv K+\phi\mbox{d}v(\phi )/\mbox{d}\phi$. While the operator algebra of Section\,III 
is available, it is difficult to find the corresponding generalization of the 
``square-root of the harmonic oscillator'' operators of Eqs.\,(\ref{C1+})--(\ref{C2+}).  

The latter were most useful, however, in order to obtain a positive definite operator $\hat P$ 
that commutes with the emergent Hamiltonian $\hat{\cal H}$ and, with this, to achieve its splitting 
into parts with positive and negative spectrum, as in Eqs.\,(\ref{Hemsplit})--(\ref{Hemsplit1}). 
The vanishing commutator here is important, since it assures that this splitting is  
invariant under evolution of the system.  

Furthermore, said operators are particularly interesting, if the resulting bounded Hamilton operator  
$\hat{\cal H}_+$ is to contain leading standard field theory terms, even though 
modified by additions as in Eq.\,(\ref{Hemplus2}), for example.
  
Following these remarks, one could try and construct such operators perturbatively, 
i.e., by deforming the operators, and include step by step increasing orders in the 
interaction $v$. 

A quite different approach might be to choose: 
\begin{equation}\label{PH}
\hat P(x)\equiv\frac{1}{{\cal M}}\hat{\cal H}^2(x)
\;\;, \end{equation} 
where ${\cal M}$ is a parameter with dimensions of energy per unit volume. 
(Note that replacing the Hamiltonian density squared with the total angular momentum density 
squared would introduce a constant with dimensions of action per unit area.) This operator 
commutes with the Hamiltonian density and will lead to a 
positive $\hat{\cal H}_+$. In fact, the resulting contributions to the Hamiltonian 
are in this case simply given by:
\begin{equation}\label{Hplusminusint} 
\hat{\cal H}_\pm =\frac{1}{4{\cal M}}\int\mbox{d}^3x\;\big ({\cal M}\pm\hat{\cal H}(x)\big )^2 
\;\;. \end{equation}   

Now, imposing the constraint, $\hat{\cal H}_-\Psi =0$, one finds that on physical 
states the bounded Hamilton operator gives: 
\begin{equation}\label{Hplusint}
\hat{\cal H}_+\Psi =\frac{1}{{\cal M}}\int\mbox{d}^3x\;\hat{\cal H}^2(x)\;\Psi 
={\cal M\cdot V}\;\Psi 
\;\;, \end{equation} 
with ${\cal V}\equiv\int\mbox{d}^3x$. A surprisingly restrictive result. 

To be sure, if one wants to connect the Hamilton operator $\hat{\cal H}_+$ of Eq.\,(\ref{Hplusint}) 
to familiar quantum field theories, the difficult task of finding ``square-root of the harmonic oscillator'' 
operators reappears. Here one has to find an underlying classical model for which the emergent 
Hamiltonian $\hat{\cal H}$, cf. Eqs.\,(\ref{Hem})--(\ref{Hem1}), contains terms which are linear in 
such operators. 

%%%%%%%%%%%%%%%%%%%%%
\section{Conclusions}
The work presented here touches a number of conceptual issues 
surrounding quantum theory. 
The interpretation of the measurement process and of the ``collapse of the wave function'', in particular, 
must figure prominently in this context, together with the ``quantum indeterminism'' and the wider philosophical 
implications of the algorithmic rules comprising quantum theory as a whole \cite{BZW}. It is left for future 
studies to find out, how a deterministic framework, such as further elaborated here, allows to see them 
in a new light.  

{\it Deterministic models} which simultaneously and consistently 
can be described as quantum mechanical ones 
present a challenge to common wisdom concerning the meaning, foundations, and limitations 
{\it of quantum theory}. 
Main aspects of the present work on such a model taken from field theory can 
be summarized as follows. 

A fairly standard description of the dynamics in phase space and its 
conversion to an operators-in-Hilbert-space formalism \`a la Koopman and von\,Neumann \cite{KN} 
yield a wave functional equation which is surprisingly similar to the functional Schr\"odinger 
equation of quantum field theory. However, the emergent ``Hamilton operator'' of this 
picture, generically, lacks a groundstate, which corresponds to the spectrum not being bounded from 
below. In order to arrive at a proper quantum theory with a stable groundstate, parts 
of the Hilbert space have to be removed by a {\it positivity constraint} which is preserved by the Hamiltonian 
flow. 

In the present example, this has been discussed based on simple supersymmetry properties 
of the underlying classical model. The important role of {\it ``square-root of the harmonic oscillator''} 
operators in constructing the constraint operator has been pointed out, and they have been constructed 
in the limit of classically 
noninteracting scalar and fermionic fields, the latter being represented by nilpotent Grassmann valued
variables. Several comments on the interacting case have been made, where they may   
be constructed in perturbation theory. In particular, these operators 
promise to be important in emergent quantum models that smoothly connect to standard field theories with 
leading quadratic kinetic energy terms.    

Here I should like to conclude with a more speculative remark concerning the dynamical 
{\it origin of the positivity constraint}, which has been introduced and interpreted as a ``loss of information'' 
at the fundamental dynamical level earlier \cite{tHooft01,Vitiello01,Blasone04}. The latter 
anticipates a still unknown, possibly dissipative information loss mechanism in the  
classical theory beneath, such as due to an unavoidable coarse-graining in the description 
of some deterministic chaotic dynamics. This would turn the system under study into 
an open system.  

However, the discussion in Section\,IV indicates a complementary point of view. There is 
a great deal of symmetry between the operators $\hat{\cal H}_+$ and $\hat{\cal H}_-$ which 
are responsible for the evolution of the system as well as for the selection of the physical states.  
In fact, since the emergent functional wave equation is linear in 
the time derivative, 
positive and negative parts of the spectrum of the emergent Hamiltonian $\hat{\cal H}$, see 
Eqs.\,(\ref{Hem})--(\ref{Hem1}), can be turned into each other by reversing the direction 
of time. Correspondingly, the roles of $\hat{\cal H}_+$ and $\hat{\cal H}_-$ can be 
exchanged. 

This suggests that giving preference to one over the other in   
determining the physical states  
may be a contingent property of the system. It typically occurs in situations where a symmetry 
is dynamically broken.  

Let us consider an extension of the present model which schematically incorporates  
such an effect. Introducing a local ``order parameter'' $\hat O$, 
take the new Hamilton operator density: 
\begin{equation}\label{Hemsplitsym} 
\hat{\cal H}_\ast (x)\equiv
\hat{\cal H}_+(x)-\hat{\cal H}_-(x)\mbox{tanh}\hat O(x)
\;\;, \end{equation} 
with $[\hat{\cal H}_\pm (x),\hat O(x')]=0$ and, for example,  
$\hat O\equiv (\hat P-{\cal M})/{\cal M}$ or $\hat O\equiv (\hat{\cal H}^2-{\cal M}^2)/{\cal M}^2$. 
The positive operators $\hat{\cal H}_\pm$ are as defined in Eqs.\,(\ref{Hemsplit})--(\ref{Hemsplit1}), 
$\hat P$ is positive definite, cf. Section\,IV, 
and ${\cal M}$ denotes an energy density parameter. All operators here commute.     

Therefore, the eigenstates of $\hat{\cal H}_\ast$ can be separated into two complementary sets, 
$\{\Psi_+\}$ and $\{\Psi_-\}$,   
with $\hat{\cal H}_-\Psi_+=0$ and $\hat{\cal H}_+\Psi_-=0$, respectively. Furthermore, 
they can be ordered according to the eigenvalues of $\hat{\cal H}=\hat{\cal H}_+-\hat{\cal H}_-$ or $\hat P$.   

For large values of the order parameter, at high energy, loosely speaking, the symmetry 
is restored and asymptotically $\hat{\cal H}_\ast\approx\hat{\cal H}_+-\hat{\cal H}_-$. In this regime, 
the system behaves classically, corresponding to an emergent Hamilton operator with 
unbounded spectrum. Here, the role of $\hat{\cal H}_+$ and $\hat{\cal H}_-$ could  
approximately be interchanged by changing the direction of time. 

Conversely, for small values of the order parameter, 
one qualitatively finds 
$\hat{\cal H}_\ast\approx\hat{\cal H}_++\hat{\cal H}_-\mbox{tanh}(1)\ge 0$. 
This result should be compared with Eqs.\,(\ref{Pfree})--(\ref{Hemmin}), for example, 
and particularly with Eq.\,(\ref{Hemplus2}).  
Here the spectrum of $\hat{\cal H}_\ast$ is bounded from below and the system behaves quantum mechanically. 
The backbending of the negative branch of the spectrum to positive values has 
replaced the imposition of the positivity constraint, Eq.\,(\ref{constraint}).  

The precise nature of the transition between classical and quantum regimes, which is 
regulated by the parameter ${\cal M}$, depends on how and which order parameter comes into play.  
Due to its nonlinearity, which introduces higher order functional derivatives, 
it modifies the underlying phase space dynamics, see Eqs.\,(\ref{Liouville})--(\ref{Hem}). 
It will be interesting to further study such corrections, which must contribute as    
additional force terms, depending on higher powers of field momentum, for example, 
to the classical Liouville operator.  

Differently from a possible ``loss of information'' mechanism, presently all operators involved 
are Hermitian and closely related to the symmetry properties of the system.   

Such a symmetry breaking mechanism might be responsible for 
the emergent quantization also in other cases than the (pseudo)classical field theory presented here. 
Besides this, 
models that incorporate interacting fermions and gauge fields are an important topic 
for future study. Furthermore, time reparametrization or general diffeomorphism invariance 
should naturally be most interesting to consider in the framework of deterministic quantum models. 

%%%%%%%%%%%%%%%%%%%%%%%%%%%
\section*{Acknowledgements} 
I wish to thank A.\,DiGiacomo for many discussions and  
kind hospitality at the Dipartimento di Fisica in Pisa and  
G.\,Vitiello, M.\,Blasone, L.\,Lusanna, E.\,Sorace, and M.\,Ciafaloni for helpful remarks and   
discussions during visits at both, the 
Dipartimento di Fisica in Salerno and in Firenze. 

%%%%%%%%%%%%%%%%%%%%%%%%%%%

%%%%%%%%%%%%%%


\begin{thebibliography}{99}
\bibitem{footnote1} Closer inspection shows that the constraint considered  
in Ref.\,\cite{I05} is not sufficient to turn the classical model there   
into a quantum mechanical one. Presently, this is solved differently   
in Section\,IV.
\bibitem{footnote2} Grassmann algebras and 
analysis over supernumbers are presented in detail by DeWitt \cite{FDeW}.
\end{thebibliography}
\end{document}